\documentclass[12pt,a4paper]{article}
\usepackage[utf8x]{inputenc}
\usepackage{graphicx}
\usepackage{color}
\usepackage{epsf}

\newcommand{\px}{$\pi\!-\!\Xi$ }
\newcommand{\pp}{$\pi\!-\!\pi$ }
\newcommand{\s}{\mbox{$\sqrt{s_{NN}}$=200}~GeV}

\newcommand{\Aoo}{\mbox{$A_{00}$}}
\newcommand{\Aii}{\mbox{$A_{11}$}}
\newcommand{\AuAu}{\mbox{Au$+$Au} }
\newcommand{\ks}{\vec{k^*} }
\newcommand{\Drs}{$\Delta {r^*}$}
\newcommand{\Xs}{$\Xi^*$}

\begin{document}
\textwidth=135mm
 \textheight=200mm
\begin{center}
{\bfseries Femtoscopy with multi-strange baryons at RHIC
}
\vskip 5mm
P. Chaloupka$^\dag$ (for the STAR collaboration)
\vskip 5mm
{\small {\it $^\dag$ Nuclear Physics Institute, Academy of Sciences of the Czech Republic,\\
250 68 Rez near Prague, Czech Republic
}}
 \
\end{center}
\vskip 5mm
\centerline{\bf Abstract}
An update on femtoscopic \px\ correlations in 
\AuAu\ collisions as measured by the STAR experiment at RHIC is presented. 
Centrality dependence of Gaussian radii and relative emission asymmetry 
was extracted for the first time.

\vskip 8mm
\section{Introduction}

The matter created in collisions of heavy ions
exhibits properties suggesting that a state with deconfined
partonic degrees of freedom was reached \cite{STAR_WhitePaper}.
%tahle veta by sla vyhodit
Current data  on spectra and elliptic flow from \AuAu collisions
at RHIC energies demonstrate that hot and dense system created in the collision
builds up substantial collectivity leading to rapid transverse expansion.

%Moreover, properties of the induced flow strongly depend on whether the
%collectivity is achieved at partonic or hadronic level.
%It's clear that multi-strange baryons, including $\Xi$,
%obtain throughout the evolution of the system substantial
%elliptic flow \cite{MultStrange_flow_Y4} comparable in magnitude
%to other particle species.
%Because of their presumably small hadronic
%cross-section multi-strange baryons are expected to undergo few
%interactions in the hadronic phase \cite{Nu_decoupling},
%and hence provide more direct probe into the early partonic stage.
A study of production of multi-strange baryons is  of high importance
since the collective behavior of these particles, manifested by large values of observed
elliptic flow together with its observed constituent quark scaling,
suggests that collective motion was already achieved prior to hadronization-
already at the partonic level.
The space-time structure of the particle emitting source is strongly affected by a collective expansion.
It shows up as an effective decrease of measured HBT radii and
a difference between average emission points for particle species with non-equal
masses.
The non-identical particle correlations can be hence used as
an independent cross-check of flow of multi-strange baryons in heavy-ion collisions
by measuring the flow-induced emission asymmetry.

%It was already shown that in heavy-ion collisions
%average emission points of pions,  kaons, and protons
%are not the same \cite{Kisiel_pion_kaon}.
Hydrodynamics-based models \cite{blastwave} predict a mass-ordering of
average space-time emission  points with the effect increasing
with a mass difference of the measured particle pair.
Since this effect is predicted to increase with a mass difference between
the particles, studying correlations in system, such as \px,
where the mass difference is large, should provide important
test of transverse expansion of the matter and flow
of multi-strange baryons.

\vskip 8mm
\section{Data selection and analysis technique}
The result presented in this paper expand our previous analysis in \cite{Chaloupka_QM06}.
In this analysis data set of \AuAu collisions at energy \s\
recorded by the STAR experiment during a Run~IV(2004) has been used.
%All the data were taken with a maximum magnetic field of 0.5~Tesla
%generated by the main STAR magnet.
The available statistics  dictates the use of only three centrality bins
corresponding to a fraction of total hadronic cross section of
0-10\%, 10-40\%, and 40-80\%.

STAR detector is capable to detect charged hadrons 
at mid-rapidity $|y|<0.8$  at full azimuthal angle
and identify primary pions  via
energy loss ({\it dE/dx}) in the STAR main TPC.
This method limits the pion transverse momenta to
\mbox{$0.08 < p_t < 0.6$ GeV/$c$}.
Primary $\Xi\ (\bar{\Xi})$-hyperons
are topologically reconstructed via their dominant decay chain
$\Xi\rightarrow\Lambda+\pi$,
$\Lambda\rightarrow\pi+p$.
The particle identification method together with acceptance of the STAR detector 
allows to reconstruct $\Xi$s at mid-rapidity in the  range of  \mbox{$0.7 < p_t < 3.0$ GeV/$c$}.

%Technique of event mixing  was used to obtain the \px
%correlation function $C(\vec{k^*})$, where $\vec{k^*}=\vec{p}_{\pi}=-\vec{p}_{\Xi}$
%denotes three-momentum of the first particle in the rest frame of the pair.
%To remove correlations of non-femtoscopic origin the mixed pairs were 
%constructed from events with sufficient proximity
%in primary vertex position along the beam direction, multiplicity
%and event plane orientation variables. 
%Pair cuts were used to remove effects of track splitting and merging.

The correlation function, obtained by event-mixing method, was corrected for 
purity of \px pairs calculated as a product of purities of both particle species.
While $\Xi$-purity was obtained from reconstructed $\Xi$
invariant mass plot as a function of transverse momentum,
the purity of pion sample was estimated from $\sqrt{\lambda}$
of the standard parametrization of the identical \pp correlation
function\cite{Lisa_Pratt_review,mercedesHBT}.
The purity correction is performed individually for each 
$\vec{k^*} = (k^*,\cos\theta,\varphi)$ bin of 
the \mbox{3-dimensional} correlation function $C(\vec{k^*})$.

\vskip 8mm
\section{Correlation functions}

\begin{figure}[htb]
\begin{minipage}[h]{0.49\textwidth}
\centering
\includegraphics[width=\textwidth,clip]{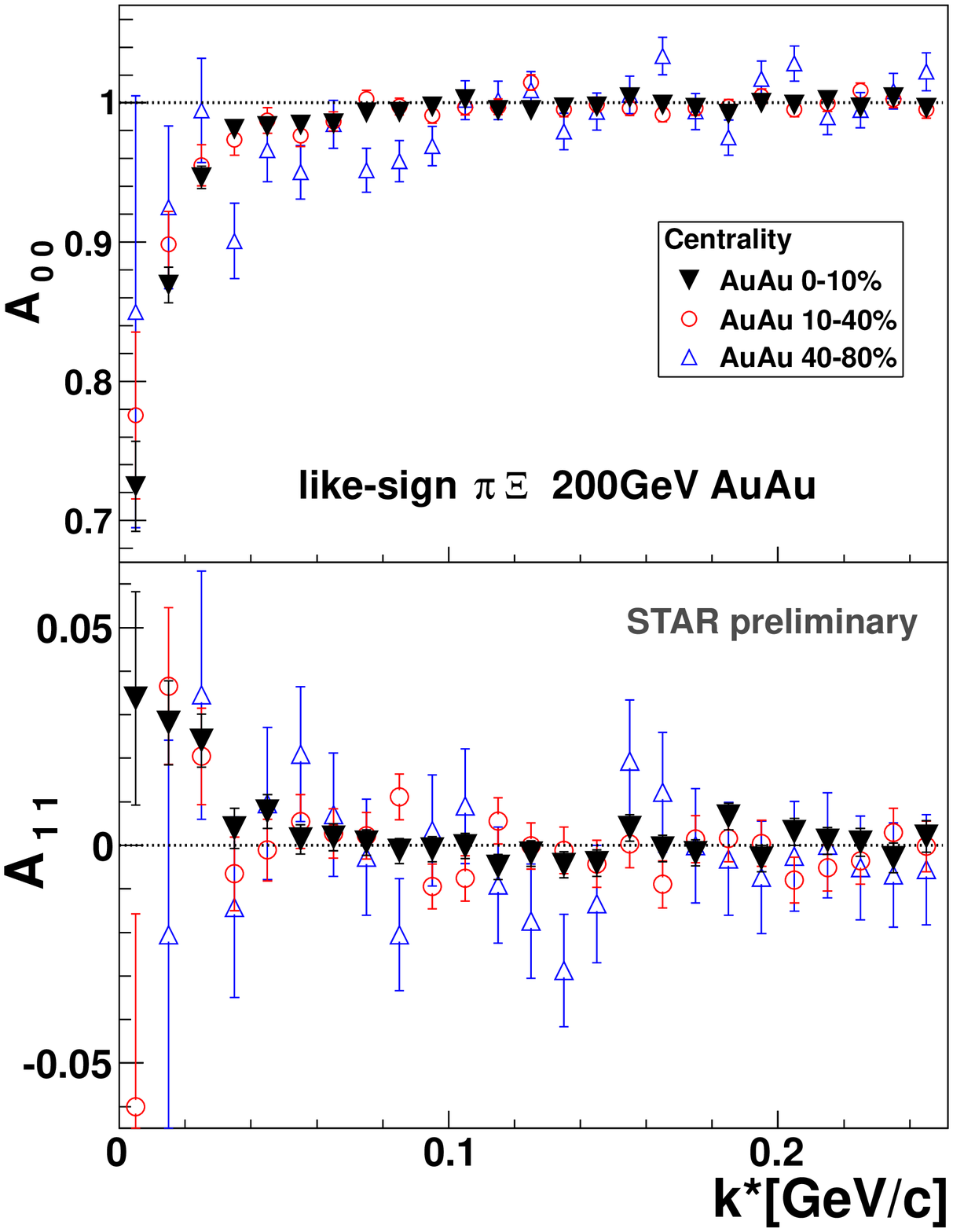}
\vspace{-0.1cm}
(a)
\end{minipage}
\hfill
\begin{minipage}[h]{0.49\textwidth}
\centering
\includegraphics[width=\textwidth,clip]{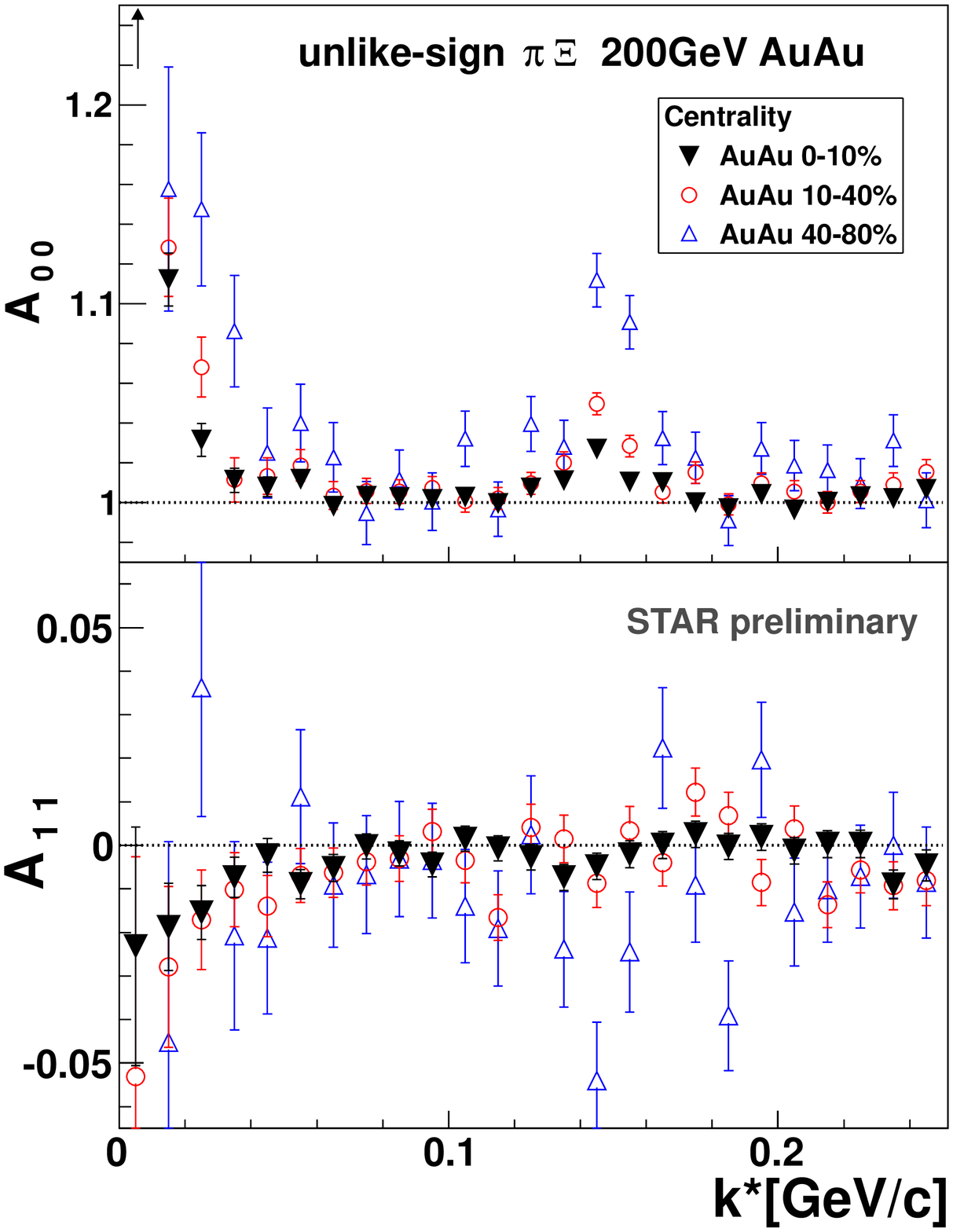}
\vspace{-0.1cm}
(b)
\end{minipage}\hfill
\caption{
	Centrality dependence of $A_{0,0}(k^*)$ and $A_{1,1}(k^*)$ coefficients of
        spherical decomposition 
        of the $C(k^*)$ from Au+Au collisions at \s\ for combined
	(a) $\pi^+\!-\!\Xi^+$ and  $\pi^-\!-\!\Xi^-$ (b) $\pi^+\!-\!\Xi^-$ and  $\pi^+\!-\!\Xi^-$  pairs.
	    }
\label{fig:SH}
\vspace{-0.25cm}
\end{figure}

\begin{figure}[htb]
  \begin{center}
    \epsfxsize=0.45\textwidth
    \epsfbox{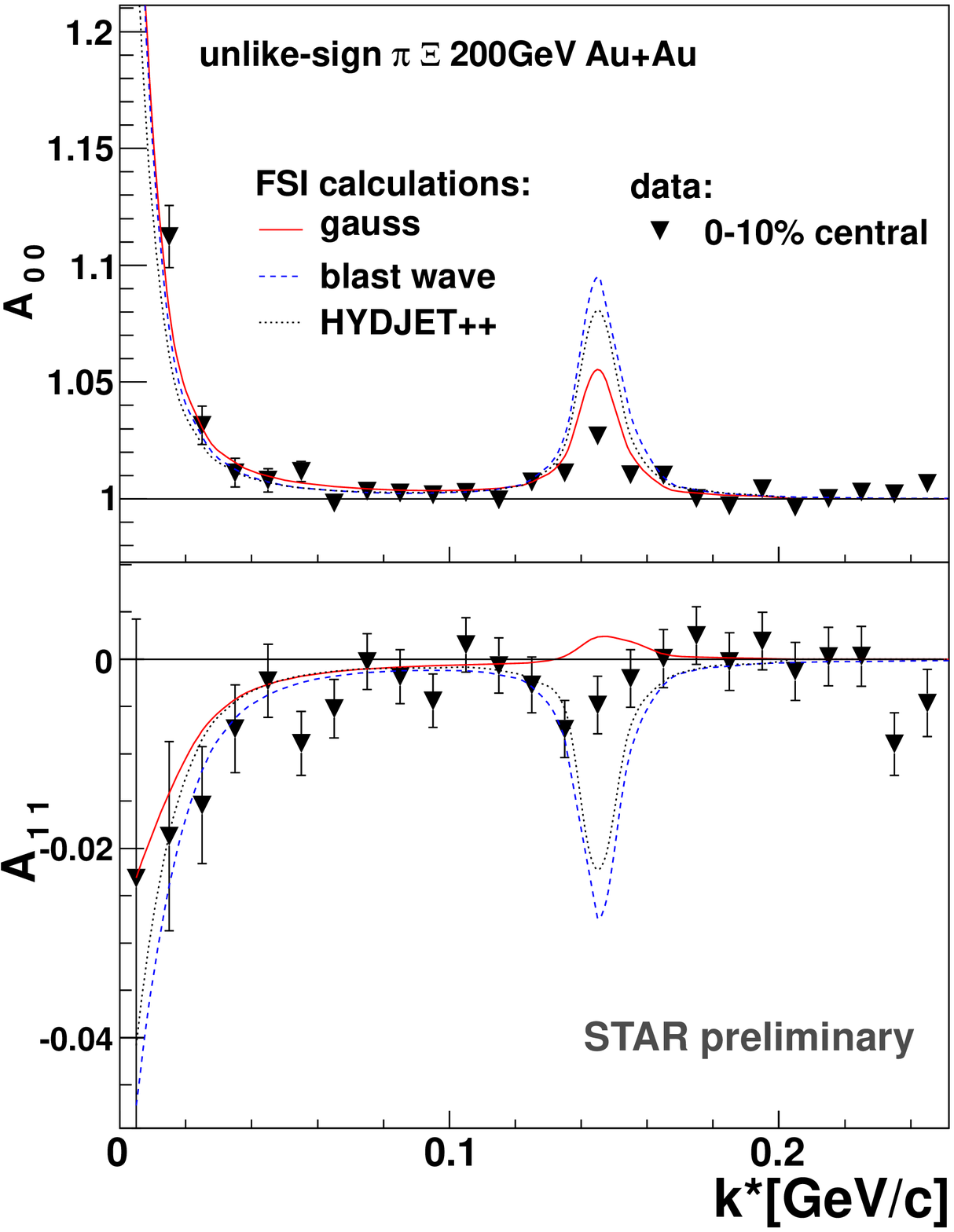}
    \caption{
        Comparison of 10\% most central data from \AuAu collisions at \s\ 
	for unlike-sign \px pairs with theoretical FSI calculations 
        using emission coordinates from Gaussian (solid)
        blast wave (dashed), and HYDJET++ (dotted) model.}
    \label{fig:FSI}
  \end{center}
\end{figure}

The 3-D correlation function $C(\ks)$ contains both, the information 
about the size of the source and the information about relative emission asymmetry.
%between the two particle species.
To access this information a decomposition  of $C(\ks)$ into
the spherical harmonics is used \cite{Chajecki_Lisa_SH}.
%The coefficients from the decomposition which are of interest
%to us are \Aoo\ and \Aii.
The coefficient $A_{0,0}(k^*)$  then 
represents 
the angle-averaged $C(k^*)$
and coefficient $A_{1,1}(k^*)$  is sensitive to the
%pair 
emission asymmetry \cite{Pratt_angularSH}. 
%In particular, in a system where both particles 
%are on average emitted at the same space-time point
%$A_{1,1}(k^*)$ vanishes.

In Figure~\ref{fig:SH}(a) and \ref{fig:SH}(b) is presented a centrality dependence 
of $A_{0,0}(k^*)$ and $A_{1,1}(k^*)$ for a combined like and unlike-sign
\px correlation function in\ \s\ \AuAu collisions.  
Both the low-$k^*$ Coulomb-dominated region and 
region at $k^*\sim 158$MeV/c, dominated by $\Xi^*(1530)$ resonance
exhibit strong centrality dependence in both coefficients.
The source size as well as the relative emission asymmetry grow
with centrality of the collision.

A comparison of the most central data with theoretical predictions is shown in Figure~\ref{fig:FSI}
for three different parametrizations of the source.
In these calculations the momenta of the particles are extracted from the real data and
their emission coordinates are calculated using given model.
The strength the correlation due to the final state interaction(FSI) is obtained  
using an approach of Pratt and Petriconi~\cite{Pratt_model}.

First calculation uses Gaussian parametrization 
of the source for both particle species.
For pions the Gaussian radii are taken from \pp analysis in \cite{mercedesHBT}.
The $\Xi$ source is then assumed to be significantly smaller($R=2$fm)
and shifted by 8fm towards the edge of the source in order to 
introduce ''by hand'' the effects of the transverse expansion.
While the theoretical correlation function describes the behavior of the data in 
the Coulomb region it gives opposite sign of the \Aii\ coefficient in the
$\Xi^*$ region. 
Since the Gaussian parametrization contains no flow-induced correlation between
particles momenta and emission coordinates it may not be valid for description
of the source at higher $k^*$. 

To incorporate more realistic description of the emission from transversely 
expanding source
we utilize hydro-inspired models: the Blastwave Model \cite{blastwave}
and HYDJET++~\cite{Lokhtin:2008xi}.
The later one also includes effects of resonances and their decays.
Both of these models, as shown in  Figure~\ref{fig:FSI}, describe qualitatively not 
only the low $k^*$ Coulomb part of the correlation function, but also 
give the right sing of \Aii\ coefficient in the region of $\Xi^*$.
It is notable that the calculations in the Coulomb part
are at or slightly bellow the data, but in the $\Xi^*$ region the
calculations strongly overshoot the data in both \Aoo\ and \Aii\ coefficients.
Despite this inability to fully calculate
strength of the Coulomb and strong part of the $C(\ks)$ at the same time it demonstrates that the 
flow-induced correlation between particles momenta and emission coordinates
are necessary for qualitative description of the measured data.

\vskip 8mm
\section{Coulomb fitting}

\begin{figure}[tb]
\begin{minipage}[h]{0.49\textwidth}
\centering
\includegraphics[width=\textwidth,clip]{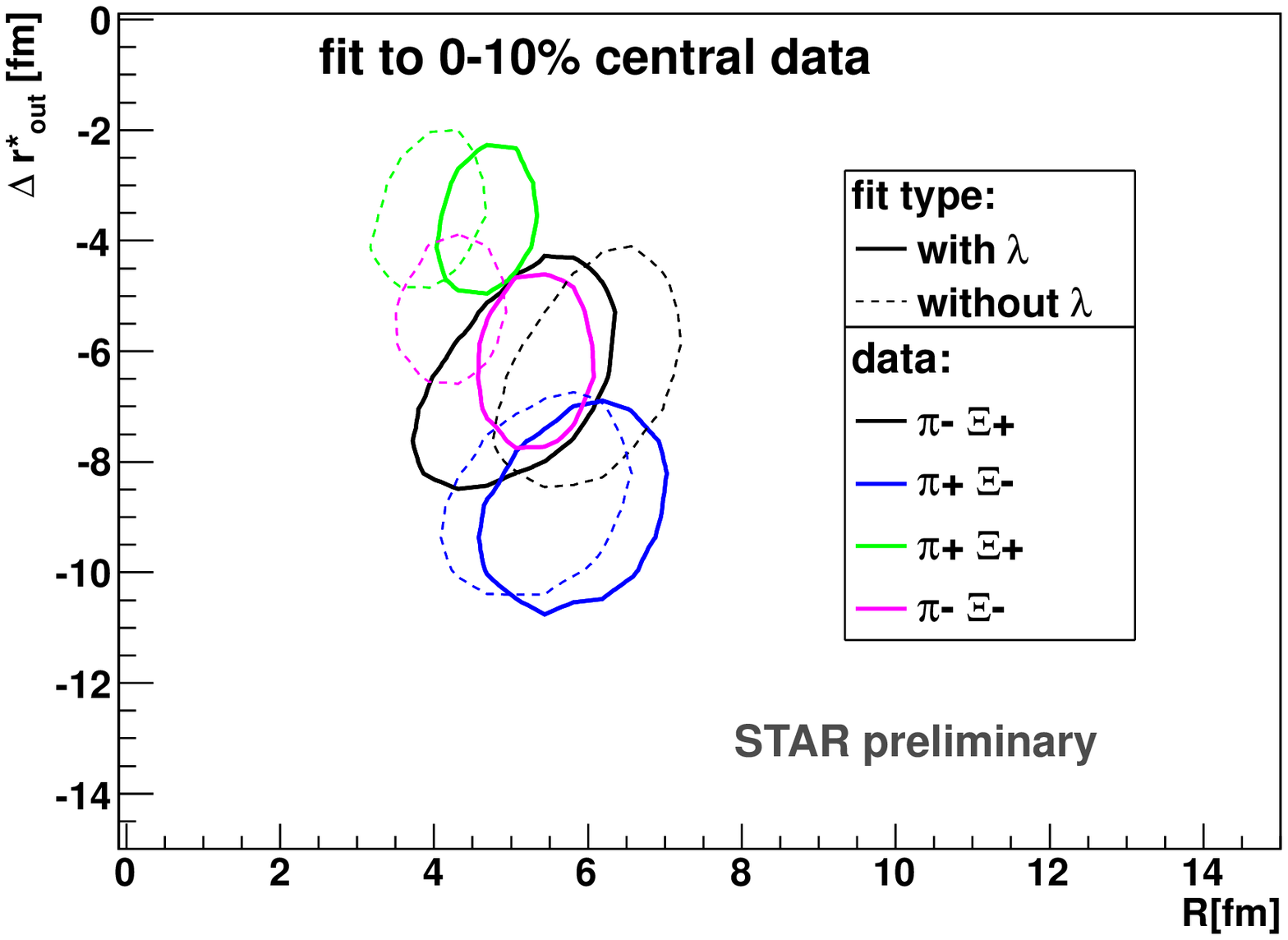}
\caption{ Results of the fit  to the
  most central 0-10\% data. 1-$\sigma$ contours in the \mbox{($R$-\Drs)} plane 
  from fits without(dashed) and with(solid) $\lambda$.
  }
\label{fig:chi}
\end{minipage}
\hfill
\begin{minipage}[h]{0.49\textwidth}
\centering
\includegraphics[width=\textwidth,clip]{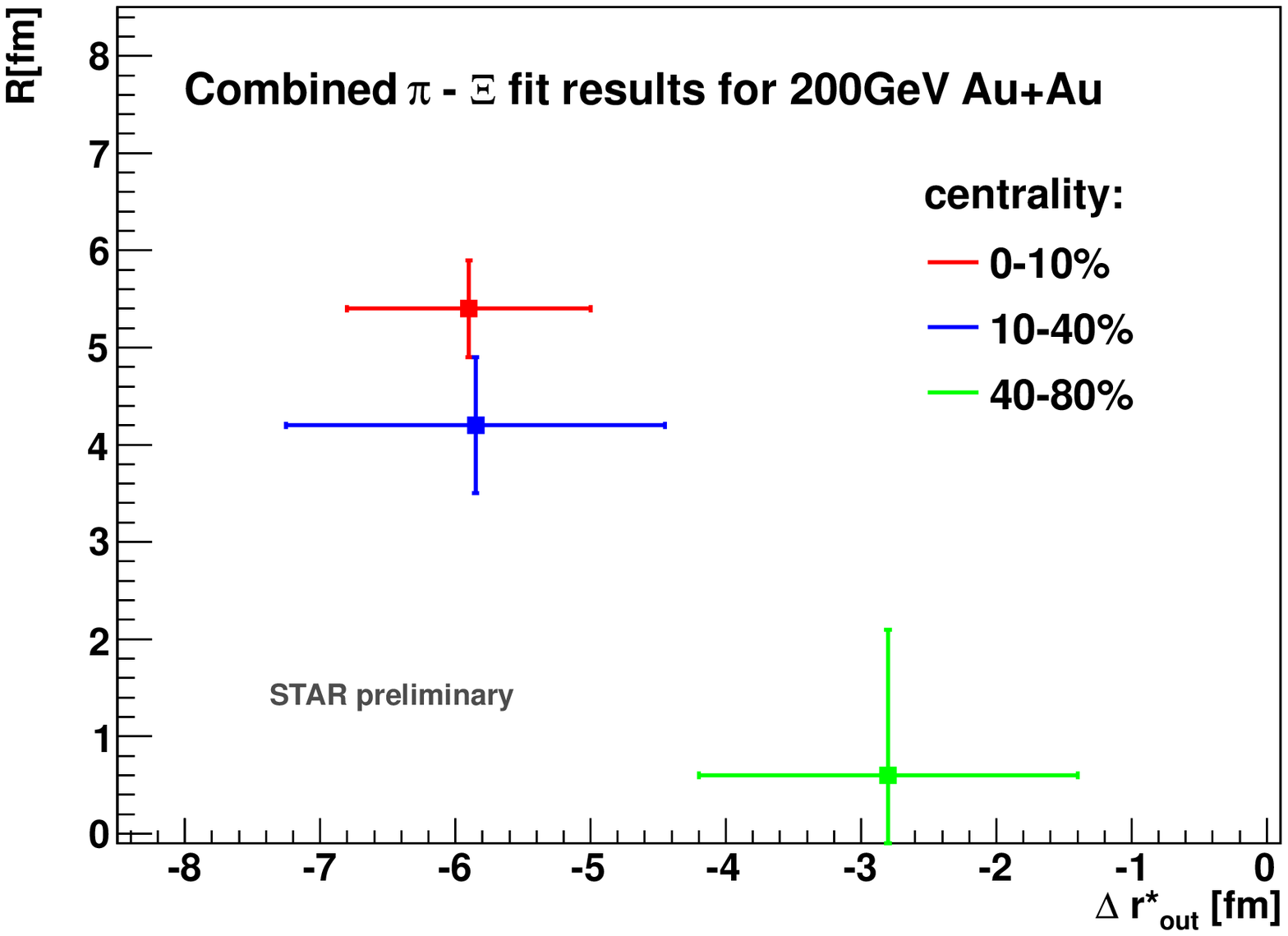}
\caption{
    Centrality dependence of the fit to the Coulomb part of the \px\ correlation function
  from \AuAu\ collisions at \s. 
}
\label{fig:FitRes}
\end{minipage}\hfill
\end{figure}

Since the Coulomb and strong region of the correlation function cannot be
described at the same time we extract the information about the size and asymmetry 
by fitting only the low-$k^*$ Coulomb region, excluding the region 
of \Xs\ resonance.
The source is parametrized by Gaussian shape with 
a single radius $R$ and a relative shift in the  {\it out}
direction \Drs:
\begin{equation}
  \label{eq:pixi_gauss}
 S_{PS}(r^{*},k^{*})\sim
\mathrm{exp}
\left[
-\left({\left(r^{*}_{out}-\Delta r^{*}_{out}\right)^2
+{r^*}^2_{side}
+{r^*}^2_{long}
}\right)/
\left({2R^{2}}\right)
\right].
\end{equation}

The fitting is done by minimizing  $\chi^2$
between calculated and real correlation function.
To stay clear of any effects of the strong interaction in the \Xs\ peak
the $\chi^2$ is calculated only from the bins with \mbox{$k^*\leq 60$~MeV/$c$}.

For most central 0-10\% event, the highest  available statistics bin, 
it is feasible to perform separate fits of each of the \px\ charge
combinations.
The fit results are presented in Figure~\ref{fig:chi} by a dashed line in the
form of \mbox{1-$\sigma$} contours in the \mbox{($R$, \Drs)} plane.
 
In the analyzes of non-identical particles
%, such as this one,
it is a standard      
procedure to first correct the the correlation function for the pair purity
and subsequently fit without the use of the $\lambda$ parameter.
Performing the purity correction in our case means to multiply
the correlation function by a factor of $\sim3$. 
Even small uncertainty in the purity factor may therefore lead to a systematic error
in values extracted from the fit,
%, especially in the Coulomb region.
For this reason we have reintroduced 
the $\lambda$ parameter in the fit of the already corrected 
correlation function in a similar way as it is done
\pp\ HBT.
In this way of fitting the obtained $\lambda$ will not be a measure
of non-purities as in~\cite{mercedesHBT}, but rather a
$k^*$-independent multiplicator of the purity corrections.
%For this reason the $\lambda$ may be, unlike in \pp\ fits, greater then 1.
%Any inaccuracy in purity correction should hence
%lead to a deviation from unity.
The convergence of the fit in the vicinity of $\lambda=1$
%can be taken as an indicator of the quality of the used
is then indicator of the quality of the performed purity correction.

In Figure~\ref{fig:chi} are shown by solid line 
results after introduction of the $\lambda$ to the fit.
In these fits $\lambda$ converges to values $0.8 \le \lambda \le 1.2$
and
%, as can be seen in the picture, 
the extracted radii, in the Figure~\ref{fig:chi}, show better
consistency between different charge combinations.
The introduction of $\lambda$ as an additional
parameter hence brought change to the extracted Gaussian radii on the order
of $10-20\%$.

The separate fitting of individual \px\ charge combinations
cannot be performed for the remaining two centralities due to low available
statistics. 
For mid-peripheral and peripheral data only 
simultaneous fit to all charge combinations
is feasible.
In the Figure~\ref{fig:FitRes} is then
presented for the first time a centrality dependence of extracted 
Gaussian radii and relative emission asymmetries from 
a simultaneous fit to the \px\ correlation function.
The results contain only statistical error of the fits.

\vskip 7mm
\section{Conclusions}
\vspace{-2mm}

An update on the \px\ femtoscopic analyzes in \AuAu\ collisions
at \s\ as measured by the STAR experiment was presented.
Comparison of measured data to 
the theoretical calculations demonstrates an importance of
flow-induced correlations between emission coordinates and momenta in 
order to qualitatively describe the correlation function in the region dominated by  
strong interaction.

We have presented a centrality dependence of extracted Gaussian
\px\ radii and relative emission asymmetries from a fit to the 
Coulomb part of the correlation function.
The obtained results show significant centrality dependence
with the values of the shift on the order of the size of the system.
This observation is in qualitative agreement with
scenarios of the evolution of the system that include significant
collective flow of multi-strange baryons.

\vskip 5mm
\leftline{\bf Acknowledgments}
This work was supported by grant LC07048 of MSMT
of the Czech Rep.

%\bibliographystyle{unsrt}
%\bibliography{bib/my.bib}

%\bibitem{a1}
%            \textit{Karsch F.  and Laermann E.} //
%            {Phys. Rev.} {D. 1994. V.50.} P.6954;\\
%            \textit{Kanaya K. } //
%            {Prog. Theor. Phys. Sup.} 1997. V.129. P.197.
\end{document}